\documentclass[prl,showpacs,twocolumn,superscriptaddress]{revtex4-1}
\usepackage{amsmath,latexsym,natbib,bm,psfrag,color,amsmath,overpic,rotating}
\usepackage{epstopdf}

\begin{document}

\title{ First-principles study of metal-induced gap states in
        metal/oxide interfaces and their relation with 
        the complex band structure}

\author{ Pablo Aguado-Puente}
\affiliation{ Departamento de Ciencias de la Tierra y
              F\'{\i}sica de la Materia Condensada, Universidad de Cantabria,
              Cantabria Campus Internacional,
              Avenida de los Castros s/n, 39005 Santander, Spain}
\author{ Javier Junquera }
\affiliation{ Departamento de Ciencias de la Tierra y
              F\'{\i}sica de la Materia Condensada, Universidad de Cantabria,
              Cantabria Campus Internacional,
              Avenida de los Castros s/n, 39005 Santander, Spain}

\date{\today}

\begin{abstract}
 We develop a simple model to compute the energy-dependent
 decay factors of metal-induced gap states in metal/insulator interfaces
 considering the collective behaviour of all the bulk complex bands
 in the gap of the insulator.
 The agreement between the penetration length obtained from the model 
 (considering only bulk properties) and full first-principles
 simulations of the interface (including explicitly the interfaces) is good.
 The influence of the electrodes and the polarization of the insulator
 is analyzed. 
 The method simplifies the process of screening materials
 to be used in Schootky barriers or 
 in the design of giant tunneling electroresistance
 and magnetoresistance devices.
\end{abstract}

% 73.40.Gk      Tunneling
% 73.30.+y      Surface double layers, Schottky barriers, and work functions 
% 77.80.Fm      Switching phenomena
% 31.15.A-      Ab initio calculations

\maketitle

 Interfaces between different oxide materials exhibit a fantastic variety
 of functional properties, some of them intrinsic to the boundary 
 between the constituent compounds~\cite{Zubko-11}. 
 A ubiquitous type of interface is that formed between a metal and an 
 insulator or semiconductor, since the presence of 
 electrodes is required for the application of electric fields in
 solid state devices.
 At any metal/insulator interface, 
 bulk Bloch states in the metal side of the junction 
 with eigenvalues below the Fermi energy and within the energy gap
 of the insulator cannot propagate into the insulating side.
 These states do not vanish right at the interface either, but they
 decay exponentially as they penetrate into
 the insulator~\cite{Heine-65}.
 Indeed, the tails of the metal Bloch states might have a 
 significant amplitude for a few layers from
 the interface, creating a continuum of gap states [the so-called
 metal-induced gap states (MIGS)]~\cite{Heine-65}.
 These MIGS are essential to determine many interfacial properties.
 The transfer of charge associated to 
 them contribute to the interfacial dipole,
 which enter into most of the models describing the formation of
 Schottky barriers.
 For this reason, theory of MIGS were deeply studied
 in semiconductor 
 heterostructures~\cite{Tung-01,Demkov-05}.
 More recently, the attention has turned to metal/oxide interfaces,
 mostly due to several works that have 
 first predicted~\cite{Zhuravlev-05,Tsymbal-06}
 and later demonstrated giant
 tunnel electroresistance~\cite{Gruverman-09,Maksymovych-09,
 Garcia-09,Burton-11}
 and magnetoresistance~\cite{Velev-05,Garcia-10,Caffrey-12}
 in ferroelectric/metal junctions,
 where the decay length of the MIGS plays a major role.

 The eigenstates of the Hamiltonian for the interface can be 
 described as the matching at the junction of
 two wave functions: an ordinary bulk Bloch state on the metal side,
 and an exponentially decaying function on the insulator side.
 Assuming that the interface is periodic in the plane parallel to the 
 boundary [referred to as the $(x,y)$ plane], then the components of the
 wave vectors parallel to the interface, $\boldsymbol{k}_\parallel$,
 are real and have to be 
 preserved when the electron crosses the junction~\cite{Zangwill}.
 Therefore, the previous matching is possible only if the two wave 
 functions do have 
 the same associated energy, symmetry, and 
 $\boldsymbol{k}_\parallel$~\cite{Bibes-11}.

 The exponential tails within the insulator decay only in the direction
 perpendicular to the junction (referred to as the $z$-direction), and can be 
 actually regarded as Bloch functions of the bulk insulator
 with an associated \emph{complex} wave vector 
 in the perpendicular direction $k_{\perp}=k_{z}+iq$~\cite{Zangwill}.
 The decay length of the wave function is given by $1/q$.
 For this reason, despite being a genuine interface property,
 characteristics of MIGS are often 
 discussed in terms of the \emph{complex band structure} (CBS) 
 of the bulk insulating material~\cite{Heine-65}. 
 This analysis is useful because it allows
 to predict the characteristics of MIGS and other interface properties from
 the simulation of a bulk material, avoiding the much more
 computationally demanding simulation of a realistic interface.
 Accordingly, many studies have taken advantage of the 
 relatively straightforward interpretation provided by the CBS
 to shed light into interface phenomena such as 
 tunneling magnetoresistance~\cite{Velev-05},
 or the confinement of two-dimensional electron gases~\cite{Janicka-09}, 
 with the discussion often focused on 
 complex bands at high symmetry points of the Brillouin zone~\cite{Velev-07}.

 Despite the previous efforts, a direct comparison of the
 spatial and energetic distribution of the evanescent states
 coming from the CBS (i.e. considering only bulk properties),
 with the MIGS obtained in a realistic simulation of the interface
 (i.e. taking into account atomic details at the junction)
 are only available in simple molecular electronic systems~\cite{Tomfohr-02}.
 Up to our knowledge, this study is absolutely absent 
 in complex oxide interfaces,
 where extra degrees of freedom such as the spontaneous polarization
 in the insulator might alter and tune their properties.
 Here, we analyze the bulk CBS of an insulator (PbTiO$_{3}$)
 and give a recipe to estimate the decay factor of the MIGS from 
 the collective contributions of all the complex bands. We show, 
 contrary to earlier assumptions, that the analysis of the full Brillouin
 zone (and not only the high-symmetry points) is necessary to accurately 
 estimate the effective decay factor of the MIGS in the insulating layer.
 We further demonstrate the validity of such approach by comparing with decay factors
 obtained from explicit simulations of realistic capacitors.
 The agreement between the penetration length estimated from the CBS
 and the one observed after simulations of realistic capacitors is very 
 good, especially remarkable when a noble metal (Pt) is used as electrode. 
 When the electrode is replaced by a metal oxide (SrRuO$_{3}$),
 the dependency of the decay factor with the
 the energy shows some extra structure due to interfacial effects 
 (symmetry filtering).
 In thin film capacitors with SrRuO$_{3}$ electrodes, 
 where the polarization of the PbTiO$_{3}$ layer is smaller than 
 the bulk spontaneous value due to imperfect screening, 
 ferroelectricity does not modify strongly the penetration length.
 Larger values of the polarization produce a decrease of the penetration of 
 the MIGS.

 We have carried out simulations within the 
 local density approximation as implemented in two different codes:
 {\sc Siesta}~\cite{Soler-02} for the calculations of realistic interfaces, 
 and {\sc Quantum-Espresso}~\cite{Giannozzi-09} 
 for the CBS computations of bulk 
 materials~\cite{Smogunov-04}. 
 The details of the simulations can be found in the supplemental data.

 \emph{Complex band structure of bulk PbTiO$_3$.}
 In order to face the difficulty of the problem step by step,
 we have proceed first with a centrosymmetric tetragonal phase ($P4/mmm$ space group),
 assuming the in-plane theoretical lattice constant, $a$, of an 
 hypothetically thick SrTiO$_{3}$ substrate and only allowing the 
 out-of-plane lattice constant, $c$, to relax.
 This is the structure that PbTiO$_{3}$  
 displays at the center of the unpolarized capacitor.

 In essence a CBS calculation
 consists in sampling the two-dimensional Brillouin zone (2DBZ) parallel to
 the interface, and for every $\boldsymbol{k}_\parallel$ 
 search for complex $k_{\perp}$ associated with real eigenvalues
 of the system.
 Imaginary bands always connect
 extrema of other bands (either real or complex)~\cite{Chang-82},
 and it is often assumed that complex bands with the shortest 
 imaginary wave vector $q$ are those connecting the edges
 of the valence and conduction bands.
 The analysis of the real band structure  
 (see Suppl. Fig. 2 in supplemental data) 
 reveals that, in the non polar P4/$mmm$ phase, PbTiO$_3$
 possesses a direct gap at X, $(\pi/a,0,0)$, 
 and a slightly larger one at Z, $(0,0,\pi/c)$.
%  Therefore, we can presume that 
%  complex bands with the shortest imaginary 
%  wave vector $q$ are going to depart from 
%  those high symmetry points.
%

 \begin{figure}[h]
    \begin{center}
       \includegraphics[width=\columnwidth]{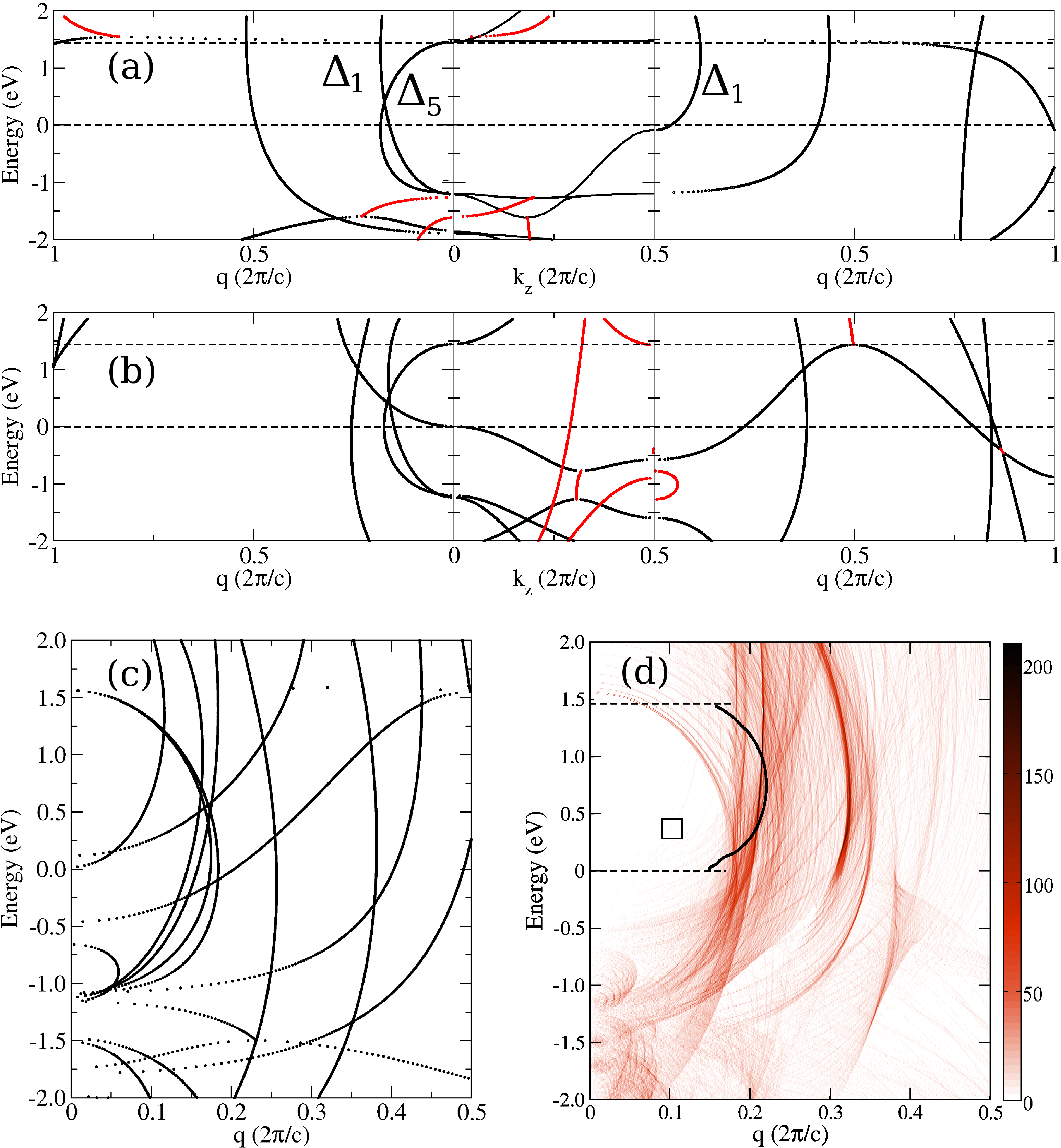}
       \vspace{-20pt}
    \end{center}
    \caption{ (color online) Complex band structure of centrosymmetric
              PbTiO$_3$ at (a) $\boldsymbol{k}_\parallel = \bar{\Gamma}$
              and (b) $\boldsymbol{k}_\parallel = \bar{{\rm X}}$. 
              Black lines at central panels correspond to bands
              with a real value of $k_\perp$. At left and right panels,
              complex bands with the form $k_\perp=iq$ and 
              $k_\perp=\pi/c+iq$ respectively, are plotted as black lines. 
              In red we plot complex bands with 
              $k_{z} \ne 0$ or $\pi/c$, with the 
              projections on the real plane plotted on the central panel
              and their projections on the imaginary plane on the
              side panels (following Ref.~\cite{Chang-82}). 
              Horizontal dashed lines delimit the band gap.
              In (c) we plot together the imaginary part of bands with
              complex values of $\boldsymbol{k}$, with 
              $\boldsymbol{k}_\parallel = \bar{\Gamma}$ or $\bar{{\rm X}}$.
              (d) Density of states (in arbitrary units) with respect 
              to the energy and the imaginary
              wave vector $q$. 
              The density of complex bands is 
              plotted in a red scale, where darker regions
              represent a larger density of bands. 
              The black solid line represents the effective penetration length
              computed from the CBS (see text).
              The zero of energies corresponds to the top of
              the valence band.
            }
    \label{fig:CBS}
 \end{figure}
 
 Complex bands at such high symmetry points 
 have an associated $\boldsymbol{k}$ 
 with the form
 $(\pi/a,0,iq)$ and $(0,0,\pi/c+iq)$ respectively, which correspond
 to complex bands at $\boldsymbol{k}_\parallel = \bar{{\rm X}}$ and
 $\bar{\Gamma}$ in the 2DBZ, shown in Fig.~\ref{fig:CBS}(a) and (b).
 We observe that the states with the shortest $q$ (largest penetration lengths)
 consist of a $\Delta_{1}$ singlet in the lower energy
 part of the band gap and a $\Delta_{5}$ doublet for energies close to the 
 conduction band edge.
 The smaller decay factor $q$ takes values from 0.10 to 0.15 
 in units of $(2\pi/c)$ at the center of the gap and
 tends to zero as the energy approaches the band edges.
 In Fig.~\ref{fig:CBS}(c) we plot together all the energy bands 
 with associated complex values of 
 $\boldsymbol{k} = 
 (\boldsymbol{k}_\parallel, k_{z} + iq)$, with 
 $\boldsymbol{k}_\parallel = \bar{\Gamma}$ or $\bar{{\rm X}}$.
 Nevertheless, we have to keep in mind that
 the CBS at $\bar{\Gamma}$ and $\bar{{\rm X}}$ might not 
 be representative of the whole CBS of the system, 
 since they are only two points of special symmetry 
 of the total 2DBZ. 
 For a comprehensive analysis, we should gather into
 a plot like Fig.~\ref{fig:CBS}(c) the imaginary part of
 complex bands coming from all $\boldsymbol{k}_\parallel$. 
 Instead of this,
 that would result into some of the infinite number of 
 complex bands obscuring others, we plot the ``density of
 complex bands'' in Fig.~\ref{fig:CBS}(d).
 In this 2D-histogram, for every
 value of the energy and imaginary part of the wave vector
 $q$, we represent in a red scale the number of complex bands contributing
 to that $q(E)$ point. 
 The analysis of Fig.~\ref{fig:CBS}(d) reveals that the total CBS is clearly
 dominated by bands at $q \sim 0.2$ and $q \sim 0.3$.
 To identify the $\boldsymbol{k}_\parallel$ that contribute
 to those clusters of bands we resolve the structure and,
 for every $\boldsymbol{k}_\parallel$, 
 we plot the smallest $q$ at a given energy 
 [see Suppl. Fig. 3(a) in the supplemental data].
 We find that from energies between the middle of the gap and the bottom
 of the conduction band, the most penetrating bands come from 
 $\boldsymbol{k}_\parallel$ lying in a narrow rectangle centered along
 the $\bar{\Gamma}-\bar{{\rm X}}$ path,
 while from the top of the valence band to the center of the gap, 
 the bands with smaller $q$ are due to states in small circular 
 (around $\bar{\Gamma}$) or square (around $\bar{{\rm X}}$) regions
 centered at the high-symmetry points. 
 The much larger area of the 2DBZ outside those sections
 contributes to the darker cluster of imaginary bands at $q \sim 0.2$.
 Similar results have been previously reported for other 
 perovskites~\cite{Velev-05,Velev-07,Hinsche-10,Wortmann-11}.
 The main messages that can be drawn might be summarized as follows:
 (i) the most penetrating bands (the most important for tunneling, 
 especially in thick films) come from high-symmetry points or lines in the 2DBZ
 whose relative weight is small. 
 That is the reason why those bands might be totally inappreciable in the
 2D-histogram of Fig.~\ref{fig:CBS}(d)
 (see region marked with a square, where the bands with $q \sim 0.1$ 
 at $\bar{{\rm X}}$ and $\bar{\Gamma}$ previously discussed were expected); and
 (ii) for ultra-thin dielectric layers or for the computation of integrated
 quantities (such as the interfacial dipole) it is important to 
 consider higher order 
 imaginary bands and non high-symmetry points in the 2DBZ, since
 associated phenomena might be
 dominated by them~\cite{Mavropoulos-00,Butler-01}. 

 It is sensible to think that the charge transferred from the metal to the 
 insulator through the MIGS reflects the collective 
 contribution of all complex band in the gap of the insulator.
 Here we also give a recipe to infer from the CBS an effective value of the
 imaginary wave vector for a given energy, 
 $q_{\rm eff}^{\rm CBS}$. 
 This can be estimated by fitting 

 \begin{equation}
    \sum_{\boldsymbol{k}_\parallel} \sum_{n} 
         e^{-2q_{n} (E, \boldsymbol{k}_\parallel) z} 
    \sim
         e^{-2q_{\rm eff}^{\rm CBS} (E) z} ,
    \label{eq:deffqeffcbs}
 \end{equation}

 \noindent which comparison with the $q_{\rm eff}$ obtained from
 the realistic simulation of the interface is discussed below.

 \emph{MIGS in first-principles simulations of capacitors.}
 The explicit influence of the 
 metal/ferroelectric interface in the decay of the MIGS 
 is captured in the simulation of realistic capacitors 
 from first-principles. 
 Here, we have simulated short-circuited 
 (SrRuO$_3$)$_{m}$/(PbTiO$_3$)$_{n}$ and (Pt)$_{l}$/(PbTiO$_3$)$_{p}$ 
 capacitors as model systems, with $m$ = 9.5, $n$ = 8.5, $l$ = 9.5
 and $p$ = 6.5 unit cells, respectively.
 In both cases the PbTiO$_3$ films are terminated in 
 a TiO$_2$ atomic layer, and we have checked that the band-alignment
 in the most stable paraelectric configuration is 
 non-pathological~\cite{Stengel-11}. 

\begin{figure}
    \begin{center}
       \includegraphics[width=\columnwidth]{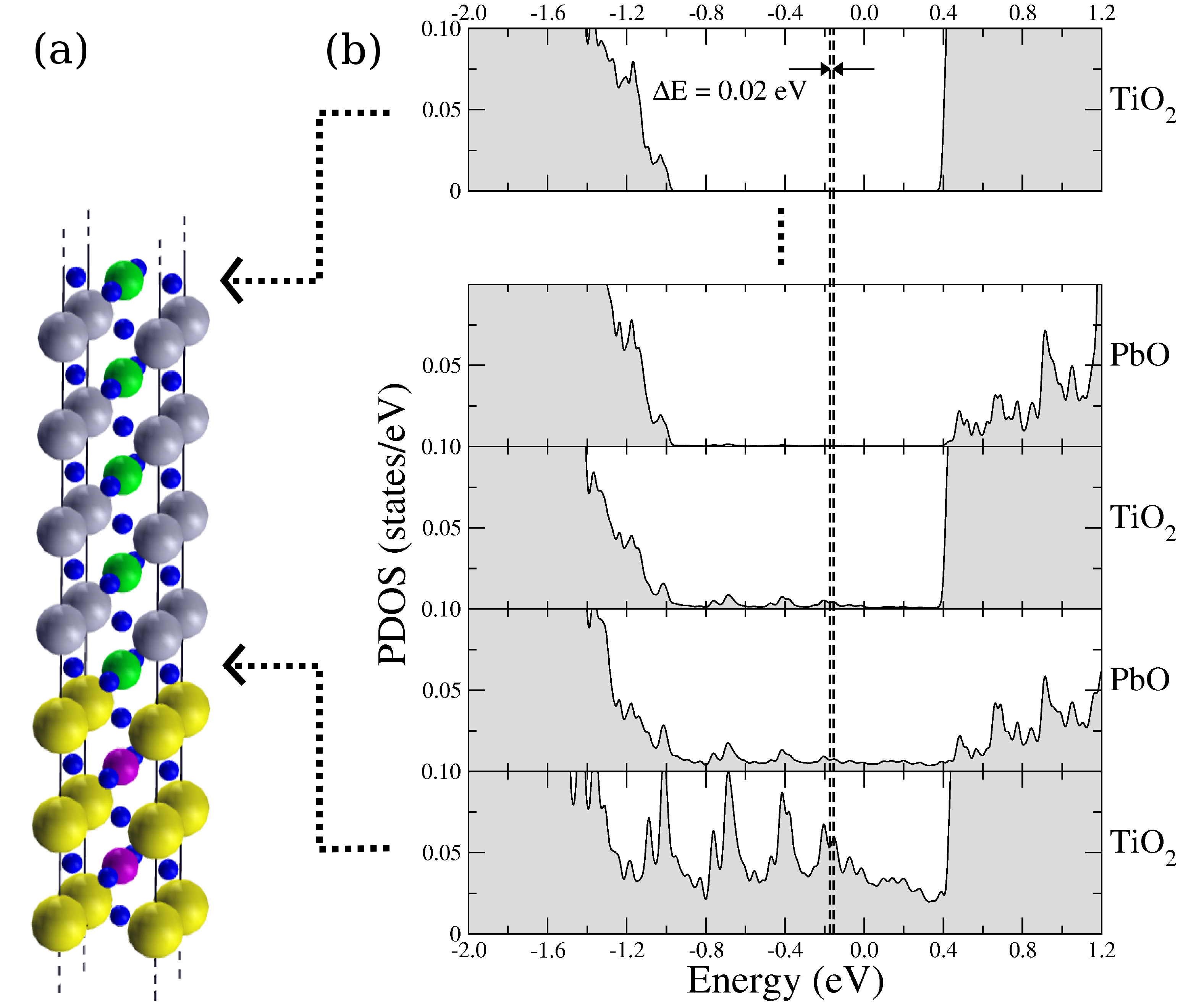}
    \end{center}
    \caption{(color online) (a) Schematic representation of the 
             interface region of the simulated
             (SrRuO$_{3}$)$_{9.5}$/(PbTiO$_{3}$)$_{8.5}$ superlattice
             in an unpolarized configuration 
             (Ti atoms in green, Ru in magenta, O in blue,
             Sr in yellow and Pb in gray). Only one of the two symmetric 
             interfaces is shown.
             (b) Layer-by-layer PDOS on the atoms at the different layers,
             as denoted by the layer labeling. The plot at the
             top corresponds to the central atomic plane of the
             PbTiO$_3$ film.
             Dashed lines delimit one of the energy windows used 
             to perform the integrations
             in Eq. (\ref{eq:Qs}).
            }
    \label{fig:PDOS}
 \end{figure}

 The analysis of MIGS in full capacitor simulations requires 
 to work with a spatially resolved density of states. 
 Here, we have chosen the layer by layer 
 projected density of states (PDOS), 
 as defined in Eq.~(21) of Ref.~\cite{Stengel-11}.
 The results for the SrRuO$_3$/PbTiO$_{3}$ capacitor in the unpolarized
 configuration are plotted in Fig.~\ref{fig:PDOS}(b)
 (similar results are obtained for the Pt/PbTiO$_{3}$ interface). 
 The decay of the evanescent tails of the MIGS can be
 clearly observed, with the PDOS showing a clean gap for
 atomic layer further than two unit
 cells from the interface.

\begin{figure}[!t]
    \begin{center}
       \includegraphics[width=\columnwidth]{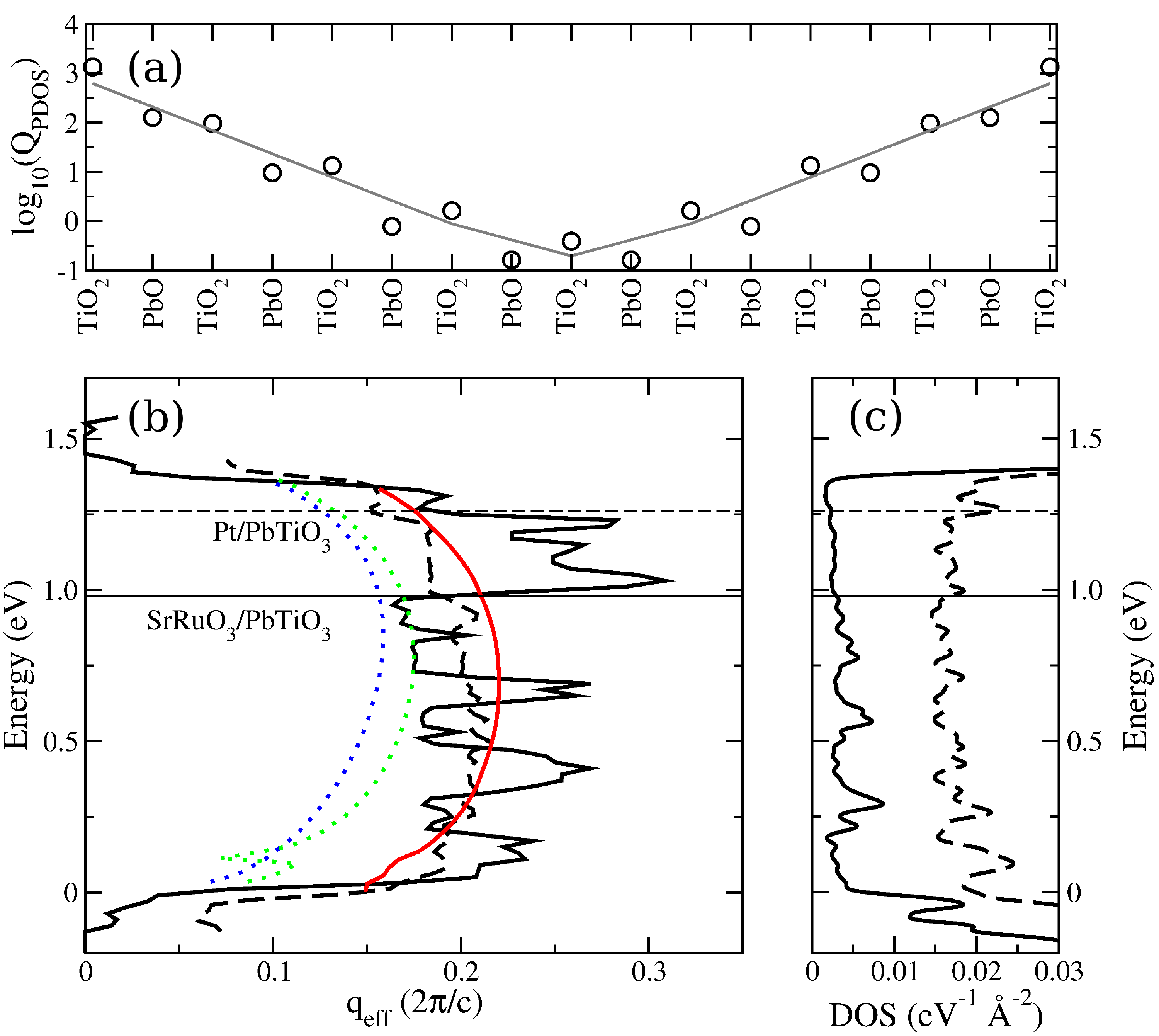}
    \end{center}
    \caption{(color online) 
             (a) Decay of the MIGS for the energy 
             window inside the gap of the SrRuO$_3$/PbTiO$_3$ capacitor
             shown in Fig.~\ref{fig:PDOS}(a). 
             The solid curve is a fit to an hyperbolic cosine, 
             from which the
             effective penetration $\delta(E)$ is obtained.
             (b) Effective imaginary wave vector $q_{\rm eff}$
             obtained from the CBS [using only bands at $\bar{\Gamma}$
             (blue dotted line), $\bar{\Gamma}$ and $\bar{X}$ (green dotted), and the full 2DBZ 
             (red line)] and from the fit of the
             decay of the PDOS inside the insulating
             material for SrRuO$_3$/PbTiO$_3$ (solid black line) and
             Pt/PbTiO$_3$ capacitors (dashed black line).
             (c) Total DOS of MIGS in the SrRuO$_3$ /PbTiO$_3$ 
             (solid black line) and Pt/PbTiO$_3$ capacitors (dashed black line)
             obtained integrating the
             layer-by-layer PDOS from the interface to the center of the
             PbTiO$_3$ film.
             In (b) and (c) the zero of energies is set to the top of
             the valence band and the horizontal lines indicate the
             Fermi level of the corresponding capacitor.
            }
    \label{fig:effq_vs_E_PDOS}
 \end{figure}

 We can obtain a measure of the spatial distribution of the probability density
 at a given energy integrating the PDOS in small windows centered
 at different energies inside the band gap,

 \begin{equation}
    Q_{\rm PDOS}(E_i,z_j) = 
      \int_{E_i-\frac{\Delta E}{2}}^{E_i+\frac{\Delta E}{2}} \rho(j,E) dE,
    \label{eq:Qs}
 \end{equation}

 \noindent where $\rho(j,E)$ denotes the PDOS on
 all the atomic orbitals of a given layer $j$, located at $z_{j}$. 
 The decay of this quantity when we move away from the interface
 provides a direct way to obtain the effective
 decay length of the PDOS, $\delta (E)$.
 Since $\delta$ refers to the decay of the \emph{density}, 
 and given the exponential 
 behaviour of the tails of the MIGS, we can relate $\delta$ 
 with an effective decay of the wave functions as
 $q_{\rm eff}^{\rm MIGS} (E) = 1/[2 \delta(E)]$.
 Numerically, $\delta (E)$ is estimated first integrating the PDOS
 after Eq. (\ref{eq:Qs})
 in energy windows of width $\Delta E = 0.02$ eV.
 Then, for every energy sampled, 
 we plot the exponentially decaying $Q_{\rm PDOS}$ curve as a function of the 
 distance to the interface [Fig.~\ref{fig:effq_vs_E_PDOS}(a)], 
 and fit it to a function 
 $Q_{\rm PDOS}(E,z) \simeq \cosh[z/\delta(E)]$.
 A hyperbolic cosine was used to account for the presence of two interfaces.
 Repeating this procedure for energy windows covering the
 whole gap we obtain the energy dependence
 of the effective imaginary wave vector, $q_{\rm eff}^{\rm MIGS}(E)$, that
 is plotted in Fig.~\ref{fig:effq_vs_E_PDOS}(b) for
 the two different capacitors. 

 \emph{Comparison of decay factors obtained from bulk and
       interface simulations.}
 In Fig.~\ref{fig:effq_vs_E_PDOS}(b) we plot the effective values of the
 decay factors obtained from a \emph{bulk} CBS calculation
 [using Eq.~(\ref{eq:deffqeffcbs}) and scaling to the band gap obtained
 with {\sc Siesta} to account for the larger gap obtained with 
 {\sc Quantum-Espresso}], and from the 
 evanescent behavior of the MIGS at the interface for PbTiO$_{3}$-based
 capacitors with SrRuO$_{3}$ and Pt electrodes.
 When the full 2DBZ is used for the calculations of $q_{\rm eff}^{\rm CBS}$ its
 agreement with $q_{\rm eff}^{\rm MIGS}$ is very good, 
 supporting the idea that the penetration of
 the MIGS is mostly determined by the bulk properties of the insulator.
 The correspondence is remarkably good for the capacitor with
 Pt electrodes, for which a featureless dependency of $q_{\rm eff}^{\rm MIGS}$
 with the energy is observed. 
 For the metal oxide SrRuO$_{3}$ a more structured shape 
 of the decay factor is noticed. 
 In this case, to understand the peaks we have to consider interfacial
 effects like the symmetry filtering: the wave functions in the electrodes
 should match evanescent states in the insulator that are compatible
 by symmetry.
 The inspection of the bulk band structure of SrRuO$_{3}$ 
 (see Suppl. Fig. 4 in the supplemental data)
 reveals the presence
 of a band below the Fermi energy with $\Delta_{5}$ symmetry along the 
 $\Gamma$-X line 
 that nicely matches complex bands of PbTiO$_{3}$ of the same symmetry,
 with $\boldsymbol{k}_\parallel$ along the 
 $\bar{\Gamma}-\bar{\rm{X}}$ line of the 2DBZ.
 Above the Fermi energy there are no bands with the appropriate symmetry
 to link the complex bands with minimal $q$.
 The matching has to be done with higher order complex bands 
 with larger $q$, 
 explaining the sharp increase in $q_{\rm eff}^{\rm MIGS}$.
 In the case of Pt, there is a band with the appropriate symmetry
 that crosses completely the band gap of PbTiO$_{3}$.
 In Fig. \ref{fig:effq_vs_E_PDOS}(b) we also plot $q_{\rm eff}^{\rm CBS}$
  obtained taking into account only complex bands from $\bar{\Gamma}$ 
  and $\bar{\rm{X}}$. As anticipated, it can be clearly noticed
  that only considering high symmetry points of the 2DBZ results in a
  severe underestimation of the effective decay factor of the MIGS.
 
 Finally, not only the decay rate of MIGS is relevant but the
 interfacial DOS of MIGS is also important. 
 We find that in this respect
 the two electrodes behave differently as well 
 [see Fig. \ref{fig:effq_vs_E_PDOS}(c)], 
 with Pt/PbTiO$_3$ capacitor 
 showing a much larger surface density of MIGS than the 
 SrRuO$_3$/PbTiO$_3$ one.
 This is consistent with the much larger DOS in the vicinity of the Fermi level
 of bulk Pt with respect to SrRuO$_3$.

 \emph{Influence of the ferroelectric polarization.}
 The influence of the ferroelectric distortion in the 
 spatial and energetic distribution of the evanescent states is known
 to play a mayor role on the electro~\cite{Velev-07,Hinsche-10,Caffrey-11}
 and magnetoresistance~\cite{Velev-05,Velev-09,Caffrey-12}
 of ferroelectric tunnel junctions.
 Unfortunately, a through analysis of such effect using the arguments 
 developed in this work is limited
 by band alignment issues~\cite{Stengel-11}
 that restrict the possible electrodes
 and the magnitude of the polarization that can be explored.
 In our case, only the ferroelectric capacitor with SrRuO$_{3}$ electrodes
 and $n$ = 8.5 is non-pathological~\cite{Stengel-11}, with a
 polarization within PbTiO$_{3}$ of 24 $\mu$C/cm$^{2}$. 
 Again, the agreement between $q_{\rm eff}^{\rm MIGS}$ (see Supplemental data 
 for details about how to compute $q_{\rm eff}^{\rm MIGS}$ in the polar state)
 and $q_{\rm eff}^{\rm CBS}$ (computed from the CBS of bulk PbTiO$_{3}$
 at the same polarization as in the capacitor) is very good
 (see Suppl. Fig. 5 in the supplemental data).
 A general increase of the effective decay factor (i.e. shorter penetration
 lengths) with the polarization is observed.
 This growth of $q$ is due to two different factors. 
 First, the polarization opens the gap.
 Second, it pushes up the flat band along the $\Gamma$-X line 
 at the bottom of the conduction band 
 in the paraelectric configuration (see Suppl. Fig. 6 in the supplmental data).
 Both facts increase the imaginary part of
 those bands that link to the bottom of
 the conduction band with $\boldsymbol{k}_\parallel$ along the
 $\bar{\Gamma}-\bar{{\rm X}}$ line. 
 This effect has already been reported for the complex bands at $\bar{\Gamma}$ 
 in BaTiO$_3$~\cite{Velev-07,Velev-09}, and
 was invoked to explain the impact of ferroelectricity in the 
 tunnel conductivity of ferroelectric junctions.

 \emph{Summary.}
 The main conclusions that can be drawn from this work are:
 (i) although the most penetrating bands are located at high-symmetry
 points, for the study of tunneling across ultra-thin layers
 or integrated quantities (where many $\boldsymbol{k}_\parallel$ contribute) 
 it is important to consider complex bands in the whole 2DBZ;
 (ii) we have given a general recipe to estimate from the CBS of the bulk
 insulator the effective decay factor 
 (and its dependence with the energy within the gap)
 of the MIGS in realistic interfaces, showing a very good agreement with the
 results obtained from first-principles simulations of capacitors; 
 (iii) to explain the fine details, especially when metal-oxide electrodes are
 used, a symmetry filtering analysis has to be performed; and
 (iv) the ferroelectric polarization can be used as a knob to increase the 
 decay factor of the CBS. 
 The accuracy of the model used to predict the effective decay length of MIGS
 would allow to screen materials to be used as barriers in tunnel 
 electro- and magnetoresistance junctions 
 based on knowledge. 
 We hope this work will encourage experiments and set the
 ground for future theoretical studies of MIGS, tunneling and related 
 phenomena.
% We hope that this work will suppose an encouragement 
% to experimentalist.

 The authors thank K. M. Rabe and M. H. Cohen for helpful discussion.
 This work was supported by the Spanish Ministery of Science and
 Innovation through the MICINN Grant FIS2009-12721-C04-02, by the
 Spanish Ministry of Education through the FPU fellowship AP2006-02958 (PAP),
 and by the European Union through the project EC-FP7,
 Grant No. CP-FP 228989-2 ``OxIDes''.
 The authors thankfully acknowledge the computer resources,
 technical expertise and assistance provided by the
 Red Espa\~nola de Supercomputaci\'on.
 Calculations were also performed at the ATC group
 of the University of Cantabria.

%%%%%%%%%%%%%%%%%%%%%%%%%%%%%%%%%%%%%%%%%%%%%%%%%%%%%%%%%%%%%%%%%%%%%%%%
%merlin.mbs 2010-03-15 4.21a (PWD, AO, DPC)
%Control: key (0)
%Control: author (8) initials jnrlst
%Control: editor formatted (1) identically to author
%Control: production of article title (-1) disabled
%Control: page (0) single
%Control: year (1) truncated
%Control: production of eprint (0) enabled
%
\end{document}

\title{ First-principles study of metal-induced gap states in 
        metal/oxide interfaces
        and their relation with the complex band structure.  
        Supplemental Material}

\author{ Pablo Aguado-Puente}
\affiliation{ Departamento de Ciencias de la Tierra y
              F\'{\i}sica de la Materia Condensada, Universidad de Cantabria,
              Cantabria Campus Internacional,
              Avenida de los Castros s/n, 39005 Santander, Spain}
\author{ Javier Junquera }
\affiliation{ Departamento de Ciencias de la Tierra y
              F\'{\i}sica de la Materia Condensada, Universidad de Cantabria,
              Cantabria Campus Internacional,
              Avenida de los Castros s/n, 39005 Santander, Spain}

\date{\today}

\maketitle

\section{Schematic representation of metal-induced gap states (MIGS).}
\label{sec:defmigs}

 \begin{figure}[h]
    \begin{center}
       \includegraphics[width=0.4\columnwidth]{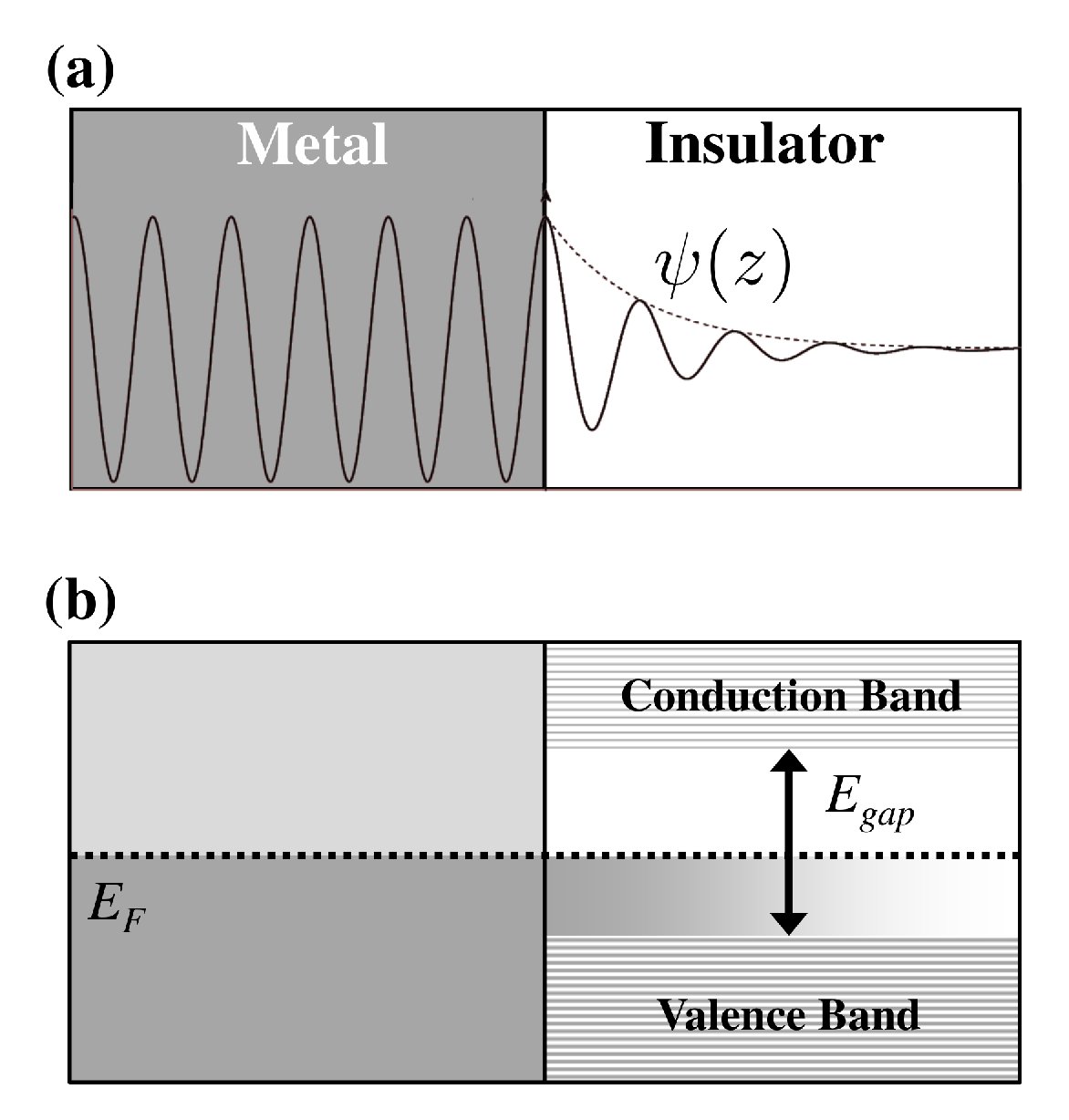}
    \end{center}
    \renewcommand{\figurename}{Suppl. FIG.}
    \caption{ Schematic representation of metal-induced gaps states in
              a metal/insulator interface. In (a), the wave function 
              of an electronic state with an eigenvalue within the insulator 
              band gap takes the form of a propagating wave function 
              in the metallic side that decays 
              exponentially on the insulating side. 
              A schematic representation of the 
              band alignment of the interface is shown in (b). 
              MIGS below the Fermi
              level of the junction are populated and the resulting 
              injection of charge is an important contribution to the
              interfacial dipole that determines the band alignment. 
           }
    \label{fig:cartoon}
 \end{figure}

\section{Computational details.}
\label{sec:methods}

 Simulations on this work have been performed within the 
 local density approximation (LDA) in the Ceperley-Alder
 parametrization~\cite{Ceperley-80} as implemented in two different codes:
 {\sc Siesta} \cite{Soler-02, siesta-web} for the
 simulation of realistic capacitors,
 and the {\sc Quantum-Espresso}~\cite{Giannozzi-09, espresso-web} 
 package for the complex band structure calculations.~\cite{Smogunov-04}
 The use of two different codes is a delicate issue
 and thus keeping strict convergence criteria becomes critical 
 in order to ensure the compatibility of all the simulations.

 In order to simulate the effect of the mechanical boundary conditions
 due to the strain imposed by the substrate, in all the simulations
 the in-plane lattice constant was fixed to the theoretical equilibrium lattice
 constant of bulk SrTiO$_{3}$ ($a$ = 3.85 \AA\ for {\sc Quantum-Espresso}
 and $a$ = 3.874 \AA\ for {\sc Siesta}).
 Since SrRuO$_{3}$ is paraelectric at room temperature, we have assumed
 non-spin polarized configurations.
 
 During the structural optimizations of both bulk PbTiO$_3$ and 
 the capacitors, a Monkhorst-Pack~\cite{Monkhorst-76,Moreno-92} mesh 
 equivalent to $12 \times 12 \times 12$ in a perovskite unit cell
 was used for the sampling of the reciprocal space.
 Tolerances for the forces and stresses were 0.01 eV/\AA\ and
 0.0001 eV/\AA$^{3}$, respectively.
 Other computational parameters, specific to each code, 
 are summarized below.

 \subsection{{\sc Quantum-Espresso}}
 Complex band structure of bulk PbTiO$_3$ has been calculated using 
 a density-functional plane-wave code with pseudopotentials.
 We have used the ultrasoft Vanderbilt pseudopotentials tabulated in the
 Vanderbilt Ultra-Soft Pseudopotential Site.~\cite{vanderbilt-uspp}
 The plane-wave cutoff was set to 40 Ry.
 For the generation of the 2D-histogram of Fig. 1(d) of the main body
 of the manuscript,
 a $48 \times 48$ mesh of $\boldsymbol{k}_\parallel$ points over
 the two-dimensional Brillouin zone (2DBZ) was sampled.

 \subsection{{\sc Siesta}}

 Computations on short-circuited Pt/PbTiO$_{3}$ and SrRuO$_{3}$/PbTiO$_{3}$
 capacitors were performed using a numerical atomic orbital method,
 as implemented in the {\sc Siesta} code.~\cite{Soler-02}
 Core electrons were replaced by fully-separable \cite{Kleinman-82}
 norm-conserving pseudopotentials, generated following the recipe given
 by Troullier and Martins.~\cite{Troullier-91}
 Further details on the pseudopotentials and basis sets
 can be found in Ref.~\onlinecite{Junquera-03.2}.

 A Fermi-Dirac distribution was chosen for the occupation of the
 one-particle Kohn-Sham electronic eigenstates, with a smearing temperature
 of 8 meV (100 K).
 The electronic density, Hartree, and exchange-correlation potentials,
 as well as the corresponding matrix elements between the basis orbitals, were
 computed on a uniform real space grid, with an
 equivalent plane-wave cutoff of 1200 Ry in the representation of the
 charge density.

 The capacitors were simulated by using
 a supercell approximation with periodic boundary
 conditions.~\cite{Payne-92}
 A $(1 \times 1)$ periodicity of the supercell 
 perpendicular to the interface is assumed. 
 This inhibits the appearance
 of ferroelectric domains and/or tiltings and rotations of the O octahedra.
 A reference ionic configuration was defined by piling up
 a number of unit cells of the perovskite oxide
 (8.5 for the SrRuO$_{3}$-based capacitor
 and 6.5 for the Pt-based capacitor), and 
 9.5 unit cells of the metal electrode.
 The PbTiO$_3$ films are always terminated in a TiO$_2$ atomic layer.

 To simulate the capacitors in a non polar configuration,
 we imposed a mirror symmetry plane at the central
 TiO$_{2}$ layer, and relaxed the resulting tetragonal
 supercells within $P4/mmm$ symmetry.
 For the ferroelectric capacitors a second minimization was carried out, 
 with the constraint of the mirror symmetry plane lifted.
 We have checked that the band-alignment
 in all systems reported here is
 non-pathological.~\cite{Stengel-11}

\section{Bulk band structure of P\lowercase{b}T\lowercase{i}O$_{3}$.}
\label{sec:bulkbandstructure}

 We compare complex band structure calculations on bulk PbTiO$_3$ 
 with properties of MIGS in capacitors where the out-of-plane cell 
 vectors were allowed to relax.
 As required by the strong sensitivity of complex bands on the 
 real band structure, bulk calculations must be performed under 
 the same symmetry constrains applied to the PbTiO$_3$ layer in the capacitor.
 It is important to note here that the bulk paraelectric phase discussed
 in this work is not the cubic phase, but a tetragonal centrosymmetric 
 $P4/mmm$ phase. The  in-plane lattice constant was fixed to the theoretical 
 one of SrTiO$_{3}$ and the out-of-plane stress was relaxed while
 the atomic positions were kept in the centrosymmetric positions.

\begin{figure}[h]
    \begin{center}
       \includegraphics[width=\columnwidth]{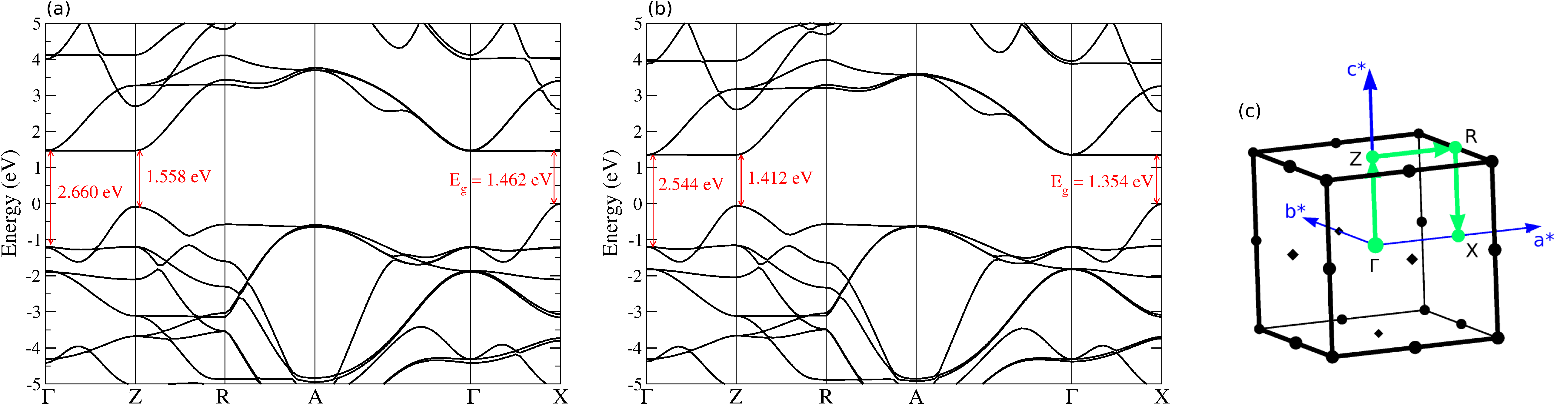}
    \end{center}
    \renewcommand{\figurename}{Suppl. FIG.}
    \caption{ (color online) Band structures of bulk non-polar PbTiO$_3$
              obtained with (a) {\sc Quantum-Espresso} and (b) {\sc Siesta}.
              A direct gap is found at X. (c) Schematic of the Brillouin zone
              in a simple tetragonal lattice.
            }
    \label{fig:bandsPTO}
 \end{figure}

 Given the sensitivity of complex bands to the size of the gap
 and the curvature of real bands they connect with,~\cite{Chang-82}
 a good agreement between the band structures obtained 
 with both codes is required to obtain comparable results.
 Suppl. Fig. \ref{fig:bandsPTO} shows that the band structures calculated 
 with the two different codes are virtually identical in shape, although the
 {\sc Quantum-Espresso} calculation displays a slightly larger gap. 
 This must be taken into account since, in principle, a larger gap should
 translate into slightly larger values of the imaginary part of the complex 
 wave vectors.  

\section{Details on the analysis of MIGS in first-principles 
         simulations of capacitors.}
\label{sec:migs_capacitor}

 The analysis of MIGS in full capacitor simulations requires 
 to work with some sort of energy-resolved probability density. 
 For this, a spatially resolved density of states is defined as

 \begin{equation}
    \rho(i, E) = \sum_{n} \int_{{\rm BZ}} d\boldsymbol{k}
                 \left| \langle i | \psi_{n \boldsymbol{k}} \rangle \right|^{2}
                 \delta (E - E_{n \boldsymbol{k}}),
    \label{eq:ldos}
 \end{equation}

 \noindent where $|i\rangle$ is a normalized function,  
 localized in space around the region of interest. 

 When $|i\rangle = |\bf{r} \rangle$ is an eigenstate of the position operator,
 the resulting $\rho({\bf r},E)$ is commonly known as {\emph local} density
 of states.
 However, strong oscillations of this function 
 due to the underlying atomic structure difficult the analysis. 
 Alternatively, the nanosmoothed version of this function might be used,
 but some interfacial properties, like precisely the decay length of the
 MIGS charge in the band gap, are sensitive to the specific convolution 
 function used for the nanosmoothing procedure.~\cite{Junquera-07}

 A reasonable choice is to work with the 
 layer-by-layer ($z$-resolved) PDOS ($E$-resolved),
 defined as in Eq.~(\ref{eq:ldos}) where
 $|i\rangle=|\phi_{nlm}\rangle$ is an atomic 
 orbital of specified quantum numbers $(n,l,m)$.
 In this case the bias of the method lies in the
 choice of the basis of atomic orbitals. 
 A sufficiently converged basis should minimize its effect since atoms of the 
 same species at different sites are equally described, so the $z$
 dependence might be considered as less biased than with previous methods.
 In Fig. 2 of the main body of the manuscript we plot the layer-by-layer 
 PDOS corresponding to the PbTiO$_3$/SrRuO$_3$ capacitor. 

 The energy distribution of the charge density converges 
 much slower with the $k$-point
 sampling than its spatial distribution.
 For this reason the PDOS was calculated performing an extra non-self
 consistent calculation with a finer $k$-point grid of 
 $54 \times 54 \times 9$.

\section{Symmetry filtering at the 
 P\lowercase{b}T\lowercase{i}O$_3$/S\lowercase{r}R\lowercase{u}O$_3$ interface}

 Fig. 3(b) of the main body of the text of this work reveals some differences 
 in the properties of MIGS of capacitors with SrRuO$_3$ or Pt electrodes.
 The dependence of the effective decay factor with the energy is essentially 
 featureless in the case of Pt electrodes, while for SrRuO$_3$ this curve 
 presents some structure. 
 The most remarkable characteristic in the latter case is a sharp increase 
 in the effective imaginary wave vector in the uppermost part of the PbTiO$_3$
 gap, right above the Fermi level of the capacitor.
 This behavior is not due to intrinsic characteristics of the complex band 
 structure of PbTiO$_3$ but to the interface. 

 The interface affects the energy dependence of the 
 MIGS penetration mainly through the symmetry filtering:
 wave functions in the electrodes should match
 evanescent states in the insulator that are compatible by symmetry,
 filtering the complex bands that contribute to the effective value.

\begin{figure}[h]
    \begin{center}
       \includegraphics[width=0.8\columnwidth]{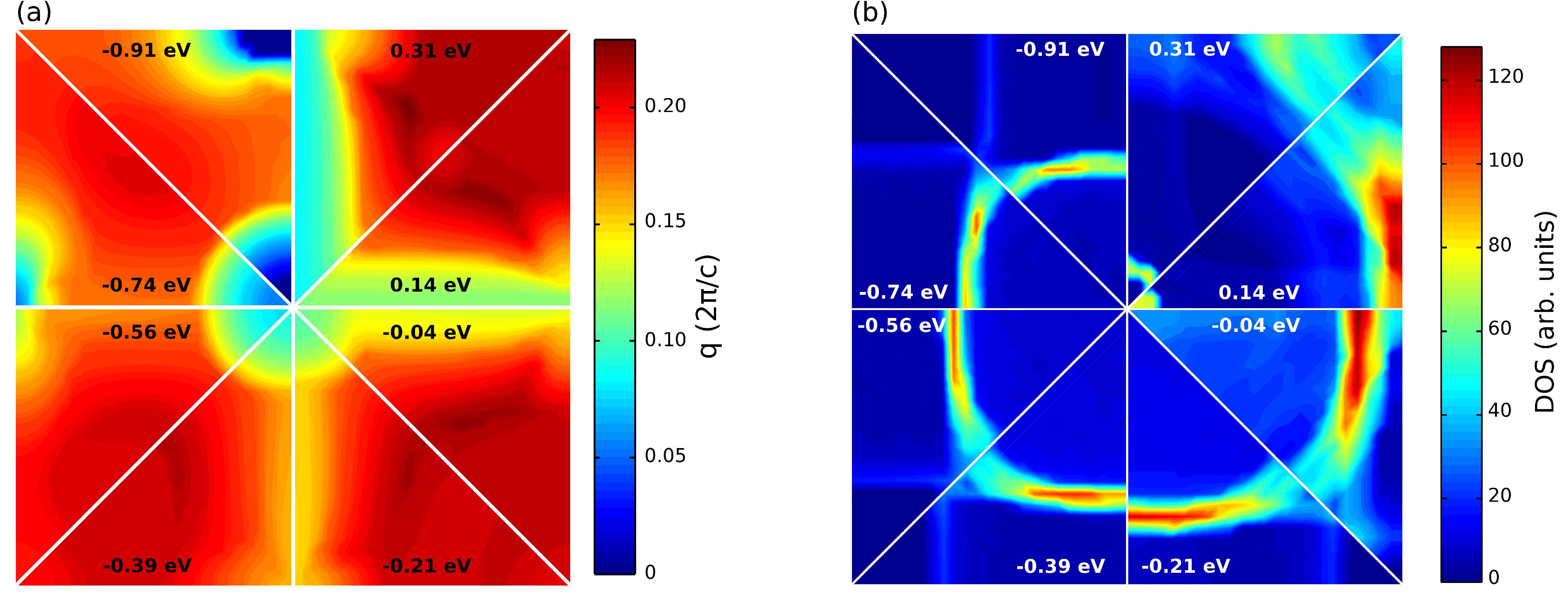}
    \end{center}
    \renewcommand{\figurename}{Suppl. FIG.}
    \caption{ (color online) (a) Minimum value of the imaginary part of the 
              complex wave vector (maximum penetration) for bulk 
              paraelectric PbTiO$_3$ over the 2DBZ. 
              (b) DOS of SrRuO$_3$ over the 2DBZ. 
              Both quantities are shown for eight different values of the
              energy, referred to the 
              Fermi level of the PbTiO$_3$/SrRuO$_3$ capacitor.
              The center of each panel represents the $\bar{\Gamma}$ point, 
              while the centers of each side symbolizes the $\bar{{\rm X}}$ 
              point.   
            }
    \label{fig:Min_q_vs_DOS}
 \end{figure}

 In Suppl. Fig.~\ref{fig:Min_q_vs_DOS}(a) we plot the complex band of minimum 
 $q$ for paraelectric PbTiO$_3$ over the 2DBZ.
 We find that from the top of the valence band to the center of the gap,
 the bands with smaller $q$ are due to states in small circular
 (around $\bar{\Gamma}$) or square (around $\bar{{\rm X}}$) regions
 centered at the high-symmetry points, 
 while from energies between the middle of the gap and the bottom
 of the conduction band, the most penetrating bands come from
 $\boldsymbol{k}_\parallel$ lying in a narrow rectangle centered along
 the $\bar{\Gamma}-\bar{{\rm X}}$ path.
As can be seen in Fig. 1 of the main body of the manuscript,
these deep-penetrating bands along the high symmetry 
$\bar{\Gamma}-\bar{{\rm X}}$ line are those that link to the 
bottom of the conduction band and have $\Delta_{5}$ symmetry
at the $\bar{\Gamma}$ point.

%  As can be seen in Fig. 1 of the main body of the manuscript,
%  the most penetrating bands at the two extreme points
%  along the high symmetry $\bar{\Gamma}-\bar{{\rm X}}$ line 
%  are of $\Delta_{5}$ symmetry for energies close to the conduction band edge.
% %
%  The same behaviour is found not only for the extreme points but
%  for the rest of $\boldsymbol{k}_\parallel$-points along the high-symmetry 
%  line.
 
 \begin{figure}[h]
    \begin{center}
       \includegraphics[width=0.8\columnwidth]{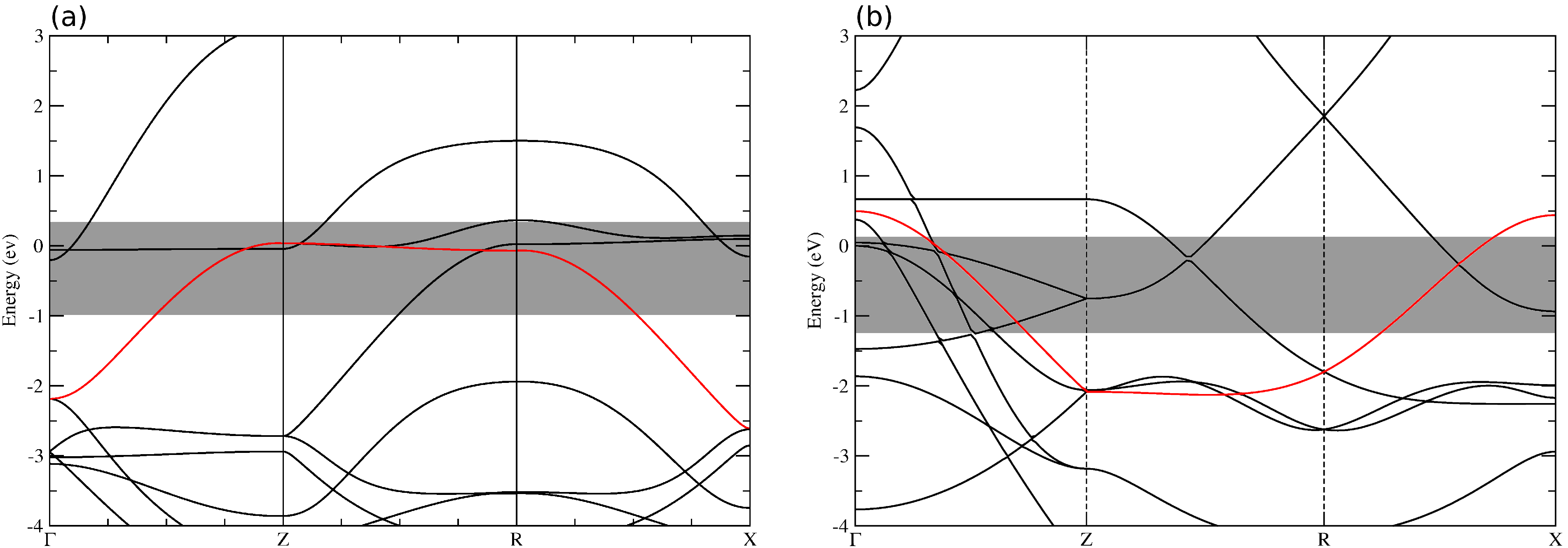}
    \end{center}
    \renewcommand{\figurename}{Suppl. FIG.}
    \caption{ (color online) Bulk real band structure of (a) bulk SrRuO$_3$ and 
              (b) bulk Pt along the relevant 
              directions in the 3-D Brillouin zone of the $P4/mmm$
              tetragonal phase.
              The bands highlighted in red has symmetry $\Delta_5$ 
              along the $\Gamma$-X line, and match
              complex bands in the PbTiO$_3$ with the same simmetry 
              with $\boldsymbol{k}_\parallel$ along
             the line $\bar{\Gamma}-\bar{{\rm X}}$. The gray area 
             indicates the band gap alignment of the PbTiO$_3$ layer
             for the corresponding capacitors.}
    \label{fig:bandasSRO}
 \end{figure}

 The symmetry analysis of the band structure of bulk SrRuO$_3$, depicted
 in Suppl. Fig. \ref{fig:bandasSRO}(a), reveals that there is a band
 (highlighted in red) with the same symmetry as the evanescent states
 discussed above all along the $\bar{\Gamma}-\bar{{\rm X}}$ line,
 and that has its maximum around the Fermi level. 
 For higher energies there are no bands with the appropriate symmetry
 to link the complex bands of minimal $q$ and form deep-penetrating MIGS.
 The matching has to be done with higher order complex bands
 with larger $q$.
 This explains the sharp increase of the effective imaginary 
 part of the wave vector for energies above the Fermi level obtained
 in the PbTiO$_3$/SrRuO$_3$ capacitor.  

 For Pt there is a band with the appropriate $\Delta_{5}$ symmetry that 
 crosses completely the band gap, 
 as shown in Suppl. Fig. \ref{fig:bandasSRO}(b).
 In other words, all the complex
 bands with minimum $q$ along the $\bar{\Gamma}-\bar{{\rm X}}$ path
 find a symmetry-compatible band in the Pt 
 electrode to link with, explaining the featureless shape 
 of $q_{\rm eff}^{\rm MIGS}$.

 Other interface-intrinsic mechanism that could affect the energy
 dependence of the $q_{\rm eff}^{\rm MIGS}$ is the 
 $\boldsymbol{k}_\parallel$-dependence of the DOS of the metal. 
 For instance, for a given energy, a DOS concentrated away from the
 high symmetry points $\bar{\Gamma}$ and $\bar{{\rm X}}$, or the
 high symmetry path $\bar{\Gamma}-\bar{{\rm X}}$
 [where the most penetrating complex 
 bands are found, see Suppl. Fig. \ref{fig:Min_q_vs_DOS}(a)]
 would cause an increase of $q_{\rm eff}^{\rm MIGS}$ at that 
 particular energy. 
 This mechanism does not seem to contribute to the sudden increase of
 $q_{\rm eff}^{\rm MIGS}$ at the Fermi level in
 the SrRuO$_3$/PbTiO$_3$ capacitor: the DOS of SrRuO$_3$ varies
 smoothly with the energy and has a significant density around the
 high symmetry points and paths near the Fermi level [see Fig. 
 \ref{fig:Min_q_vs_DOS}(b)].

 \section{Discussion about the effect of the ferroelectric polarization 
          of the P\lowercase{b}T\lowercase{i}O$_3$ layer}

 \begin{figure}
    \begin{center}
       \includegraphics[width=0.6\columnwidth]{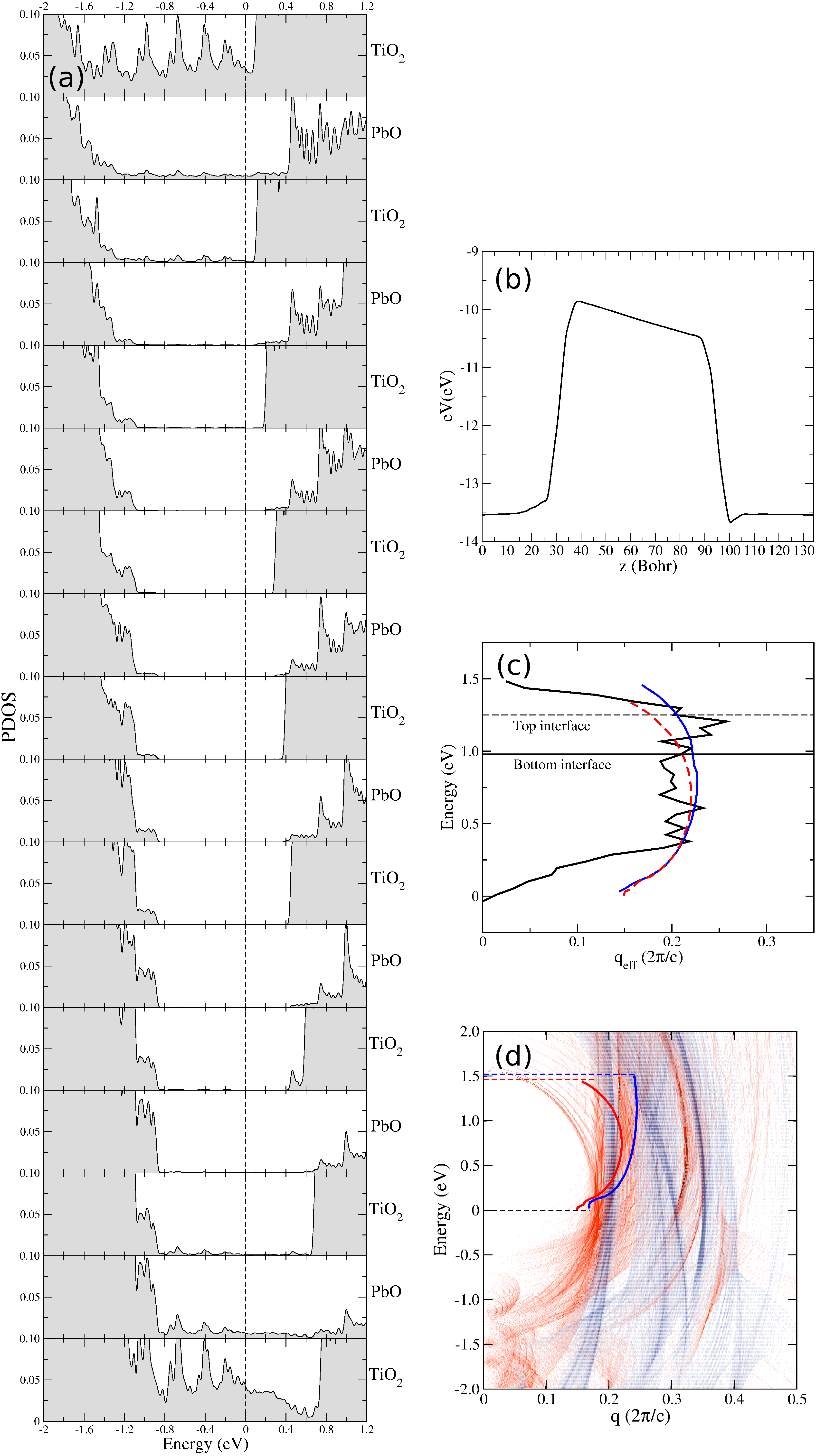}
    \end{center}
    \renewcommand{\figurename}{Suppl. FIG.}
    \caption{(color online) (a) Layer-by-layer PDOS of a polar
              [SrRuO$_3$]$_{9.5}$/[PbTiO$_3$]$_{8.5}$ capacitor. The dashed
              line indicates the Fermi level.
              (b) Profile of the macroscopically-averaged electrostatic
              potential across the capacitor.
              (c) Effective imaginary wave vectors obtained from the
              fit of the decay of the PDOS of the PbTiO$_3$ layer in the polar
              capacitor (black line), from the CBS of bulk PbTiO$_3$ 
              with the same ferroelectric distortion as in the center of the 
              capacitor (blue line) and from the CBS of centrosymmetric PbTiO$_3$
              (red line). The two horizontal lines indicate the position
              of the Fermi level at the two interfaces (note that in the capacitor
              the Fermi level is the same across the whole structure, but
              to extract $q_{\rm eff}^{\rm MIGS}(E)$ the PDOS at each atomic
              layer is shifted following the electrictrostatic potential to 
              align the band structures). 
              (d) Complex band structure of centrosymmetric P4/$mmm$
             (red) and fully relaxed ferroelectric (blue) PbTiO$_3$. 
             The plot was obtained following the same recipe as for
             Fig. 1(d) of the main manuscript. Red an blue solid lines 
             are the effective decays obtained from the CBS of
             centrosymmetric and polarized PbTiO$_3$ respectively.
             Red and blue dashed lines correspond to the bottom of the
             conduction band in each case. The zero of energies in panels (c)
             and (d) is set to the top of the valence band.
            }
    \label{fig:complex}
 \end{figure}

 Most of the discussions in the main text of the manuscript regard the 
 paraelectric phase of the PbTiO$_3$/SrRuO$_3$ capacitor. However,
 PbTiO$_3$ is a ferroelectric perovskite with a large spontaneous
 polarization in bulk \cite{ZhongW-94.1}. 
 Indeed, the influence of the polarization in the spatial
 and energetic distribution of evanescent states in 
 ferroelectric/metal interfaces has been proposed to
 contribute largely to the electro-\cite{Velev-07,Hinsche-10,Caffrey-11} 
 and magnetoresistance \cite{Velev-05,Velev-09,Caffrey-12}
 of ferroelectric tunnel junctions.
 A thorough analysis of such effect, using similar arguments as
 those developed in this paper, would be very useful in this regard.
 It should be noted, however, that the investigation of those aspects
 related with the realistic interfaces has the 
 limitation imposed by the band alignment issues reported elsewhere
 \cite{Stengel-11}, that restrict (i) the number of possible electrodes
 that can be tested within the usual approximations to 
 the exchange-correlation functional
 (only for SrRuO$_3$, a sizable polarization can be induced in the
 PbTiO$_3$ layer before causing the 
 spurious electric breakdown of the capacitor)
 and (ii) the magnitude of the 
 polarization (even for SrRuO$_3$ electrodes,
 the interface becomes pathological for polarizations 
 of the PbTiO$_3$ layer larger than 
 $\sim 40$ $\mu$C/cm$^2$ ).

 For the capacitors discussed here, we find only the 
 [SrRuO$_3$]$_{9.5}$/[PbTiO$_3$]$_{8.5}$ capacitor to display
 a correct band alignment in the polar configuration,
 with a polarization of 24 $\mu$C/cm$^2$
 after relaxing under short-circuit boundary conditions.
 Suppl. Fig. \ref{fig:complex}(a) displays the layer-by-layer
 PDOS in the PbTiO$_3$ film. The tilting of the bands due to 
 the remnant depolarizing field is clearly appreciable. 
 To extract the $q_{\rm eff}^{\rm MIGS}$ from the 
 PDOS using the same recipe developed in the main manuscript
 we have to take into account that here the band structure 
 at different atomic layers is shifted as a result of
 the depolarizing field. To perform the fitting of the decay 
 of the MIGS in the same fashion as in Fig. 3(a) of the main paper,
 we align the layer-by-layer PDOS using the macroscopic 
 average of the electrostatic potential, shown in Suppl.
 Fig. \ref{fig:complex}(b). The resulting $q_{\rm eff}^{\rm MIGS}(E)$
 for the ferroelectric configuration of the SrRuO$_3$/PbTiO$_3$
 capacitor is shown as a black line in Suppl.
 Fig. \ref{fig:complex}(c). 
 When compared with the $q_{\rm eff}^{\rm CBS}(E)$ obtained for
 bulk PbTiO$_3$ with the same ferroelectric distortion 
 a good agreement is found,
 as in the paraelectric configuration. For such
 polarization (24 $\mu$C/cm$^2$) the change in the effective decay of MIGS
 induced by the ferroelectric distortion is not very large. 
 
  \begin{figure}
    \begin{center}
       \includegraphics[width=0.45\columnwidth]{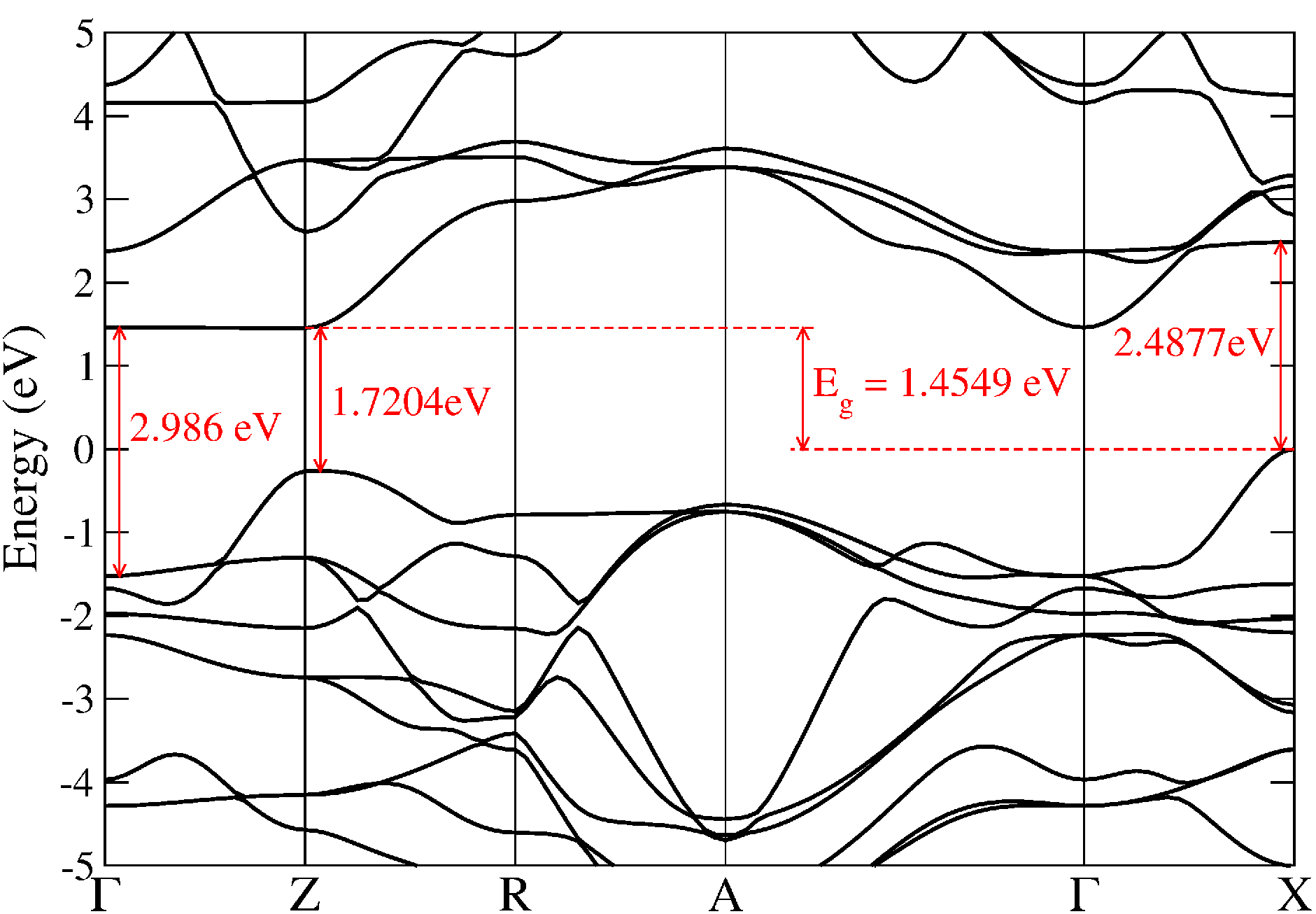}
    \end{center}
    \renewcommand{\figurename}{Suppl. FIG.}
    \caption{(color online) Band structure of bulk ferroelectric 
             PbTiO$_3$ with the in-plane
             lattice constant fixed to the theoretical one of SrTiO$_3$. Band
             structure calculated with {\sc Siesta}.
            }
    \label{fig:bands_FE}
 \end{figure}
 
 To appreciate better the effect of the ferroelectricity in the
 MIGS we compute the CBS for the fully relaxed
 bulk PbTiO$_3$ (which displays a polarization of 81 $\mu$C/cm$^2$
 when the in-plane lattice constant is fixed to that of SrTiO$_3$). 
 The resulting $q_{\rm eff}^{\rm CBS}$ is plotted 
 in Suppl. Fig. \ref{fig:complex}(d).

 The effect of the ferroelectric distortion on the band structure of 
 PbTiO$_3$ is two-folded (see Suppl. Fig. \ref{fig:bands_FE}). 
 In first place it opens the gap, 
 tending to increase the imaginary part of the complex bands
 wave vectors. In addition, it pushes up the flat band that
 constitutes the bottom of the conduction
 band along the $\Gamma$-X path in the paraelectric configuration.
 Suppl Fig. \ref{fig:complex}(d) shows that this increases greatly
 the imaginary part of 
 those bands that link to the bottom of
 the conduction band with $\boldsymbol{k}_\parallel$ along the 
 $\bar{\Gamma}-\bar{{\rm X}}$. 
 Suppl. Fig. \ref{fig:complex}(d) also shows that, while complex bands 
 at high symmetry paths of the 
 2DBZ are indeed strongly affected by the polarization of PbTiO$_3$, 
 the dense cluster of bands with $q\sim 0.2$ $(2\pi/c)$ display very little
 distortion in the polar state with respect to the non-polar one.
 The net result is 
 a decrease of the penetration of MIGS, that is larger the
 closer the energy is to the conduction band, as evidenced by the
 estimated $q_{\rm eff}^{\rm CBS}$ shown in Suppl Fig. \ref{fig:complex}(d).